\documentclass[conference]{IEEEtran}
\IEEEoverridecommandlockouts
\usepackage{cite}
\usepackage{amsmath,amssymb,amsfonts}
\usepackage{algorithmic}
\usepackage{graphicx}
\usepackage{textcomp}
\usepackage{booktabs}
\usepackage{xcolor}
\usepackage{tabularx}
\usepackage{mathtools}
\DeclarePairedDelimiter{\ceil}{\lceil}{\rceil}
\def\BibTeX{{\rm B\kern-.05em{\sc i\kern-.025em b}\kern-.08em
    T\kern-.1667em\lower.7ex\hbox{E}\kern-.125emX}}

\usepackage{soul}

\usepackage{cancel}

\begin{document}

\title{Exploring Ethereum’s Data Stores: A Cost and Performance Comparison}

\author{\IEEEauthorblockN{Periklis Kostamis, Andreas Sendros, Pavlos Efraimidis}
\IEEEauthorblockA{\textit{Electrical and Computer Engineering Dept.} \\
\textit{Democritus University of Thrace}\\
Xanthi, Greece \\
\{perikost1, asendros, pefraimi\}@ee.duth.gr}
}

\maketitle

\begin{abstract}
The cost of using a blockchain infrastructure as well as the time required to search and retrieve information from it must be considered when designing a decentralized application. In this work, we examine a comprehensive set of data management approaches for Ethereum applications and assess the associated cost in gas as well as the retrieval performance. More precisely, we analyze the storage and retrieval of various-sized data, utilizing smart contract storage. In addition, we study hybrid approaches by using IPFS and Swarm as storage platforms along with Ethereum as a timestamping proof mechanism. Such schemes are especially effective when large chunks of data have to be managed. Moreover, we present methods for low-cost data handling in Ethereum, namely the event-logs, the transaction payload, and the almost surprising exploitation of unused function arguments. Finally, we evaluate these methods on a comprehensive set of experiments.

\end{abstract}


\section{Introduction}

While the number of individuals that use blockchain systems is growing significantly, concerns about the performance and cost of using public blockchains like Ethereum\cite{b20}, discourage widespread adoption. In particular, storing data in Ethereum is expensive. However, Decentralized Applications (DApps) running on top of blockchain, usually manage some of their data on-chain.  Thereby, improper development of Smart Contracts (SCs) results in higher costs that are laid on to the users of each application. Distributed File Systems such as IPFS\cite{b24} and Swarm\cite{b22} can be used alongside Ethereum to ameliorate this problem.

According to this information we define our research question: how can data be efficiently managed in Ethereum DApps? 
In our attempt to answer this we examine a wide range of 
storage options in Ethereum as well as a hybrid on-chain/off-chain architecture. More specifically, we make a comparative study to explore the advantages and disadvantages of each solution. Regarding Ethereum, the cost of storing the data is the main criterion in evaluating our results. Also, we comment on the complexity of implementing each method and compare their efficiency concerning data retrieval. In the second case, Ethereum is used for reliable data referencing and IPFS or Swarm as the main storage. Because the cost of storing these references is fixed, we mainly focus on the performance of these platforms, in the context of uploading and retrieving data.

The structure of this paper is as follows. In Chapter 2 we discuss the related work. In the next chapter, we compare the available practices of storing data in the Ethereum blockchain and describe the main differences between IPFS and Swarm, from a theoretical point of view. In Chapter 4 we define the criteria of our experiments, outline the environment in which we conducted them and discuss our results both in terms of cost and performance. In the last chapter, we gather our conclusions and refer to future work. By making an overall analysis regarding cost and performance in the aforementioned data management methods, we aspire to provide guidance for developers of DApps.

\section{Related Work}

Examining the respective work that has been done, we believe that data storage in Ethereum is an under-researched topic. There are a few studies \cite{b6, b27, b1} that examine the gas consumption of SC but none of them research the related cost holistically by considering all possible ways of using Ethereum as a data store.

Tools have also been developed in an attempt to analyze gas cost and assist developers, but mostly, the focus is on the program structures rather than the data. Among them we singled out Gasper \cite{b10}, a cost minimization tool that analyzes bytecode to determine costly patterns in SCs. Its creators used it to analyze approximately 4000 SCs and concluded that three of these patterns were found in the majority of them. This research was continued, and new patterns have been added. There is also a work on parametric 
cost bounds in the Ethereum blockchain\cite{b4}.

Research has been done in managing data of DApps at a lower cost. However, the novel idea of storing data in a transaction’s payload is barely studied. Blockchi\cite{b8} is a tool that allows storing and retrieving JSON objects, while in \cite{b26}, a storage model is proposed for storing data from IoT sensors in hexadecimal format. In the latter, transaction references  are retained in an external database. In our work, apart from evaluating the transaction payload as a data store, we propose the following. A model of future-tracking the transaction references without the need of an off-chain tool and a way of exploiting Solidity’s ABI interface to allow working with all available data types.

In another work \cite{b9} related to our research, IoT data were stored in IPFS and Swarm while the resulting identifiers were recorded in Ethereum. Both retrieval and upload performances of the local node were studied. In 
\cite{b11}, IPFS was evaluated for remote and local retrieval latency. In our work we focus on local read-write performance of IPFS and Swarm. As far as Swarm is concerned, there is limited research regarding its performance and none that considers its recently updated protocol on which we based our benchmark. Likewise, in our experiments related to Ethereum gas consumption, we take into account the recently announced Berlin hard fork and compare the alternated storage cost model to the current one.

\section{Data Management Methods in blockchain}

\subsection{Ethereum}
Ethereum is considered to be a distributed transaction-based state machine \cite{b19}. There are two key components that enable Ethereum’s operation. The Ethereum Virtual Machine (EVM), which is a Turing-complete virtual machine with a simple stack-based architecture and the Ethereum blockchain, an append-only timestamped data structure where transactions and relevant information are stored. Each node that participates in the Ethereum network runs an instance of the EVM and holds a copy of the blockchain. State transitions are caused by the execution of transactions, which may result either in a transmission of value between accounts or an execution of EVM code associated with an account. Such code, known as SC, is essentially an immutable computer program that runs on top of the Ethereum protocol. Typically, SCs are written in high level programming languages, primarily in Solidity \cite{b13}, but must be compiled to a series of bytecode instructions in order to be executed by the EVM.

Every computation during a transaction execution is subject to a fee. 
This fee is predetermined \cite{b19} and paid in gas, the fundamental cost unit in Ethereum.
 Gas is purchased with ether prior to the transaction’s execution, at a certain gasPrice specified by the sender. Furthermore, to prevent the EVM from getting stuck in intentional or unintentional infinite loops, the notion of gasLimit was introduced. In the context of a transaction, gasLimit indicates the amount of gas the sender is willing to spend. Each block has an equivalent limit that bounds the total gas that can be used by the transactions included in it.

Since Ethereum aims to provide the necessary infrastructure for all kinds of applications, having a persistent memory area for the needs of the applications, is crucial. Indeed, every SC owns an autonomous memory area called storage, where data can be saved. Utilizing a contract’s storage also results in a gas fee. In particular, the cost of writing data on storage is proportional to the data’s size and especially high, as a means to keep the distributed database as small as possible. Consequently, storage space within the Etherenum network is a valuable resource and it should be used sparingly to store what is required for the SC's proper operation. Obviously, this limits drastically the amount of data an application can store on Ethereum.
To further minimize pointless occupation of storage space, a gas refund is given when clearing an entry, i.e., setting a value from non-zero to zero.

Except for SC storage, other data structures within Ethereum’s architecture can be used as a form of cheaper data stores. In the following paragraphs we elaborate on these alternatives.

\paragraph{Storing Data in SC Storage}

SC storage is a non-volatile key-value store, mapping 32-byte words to 32-byte words. Programming languages used for SC developing simplify the manipulation of storage by providing an abstraction on top of the EVM. Specifically, Solidity supports a plethora of static data types as well as some dynamic types, namely dynamic arrays and mappings.

\begin{figure}[htbp]
\centerline{\includegraphics[width=9cm,height=1cm]{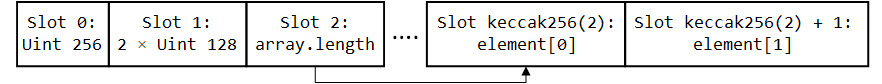}}
\caption{Layout of dynamic arrays in SC storage.}
\label{fig:arrays}
\end{figure}

Static types are stored consecutively in storage, occupying one or more slots, depending on their size. Due to their unpredictable size, dynamic data types cannot adhere to the same principles. Regarding dynamic arrays, their starting position is determined using the keccack hash function. At first, a storage slot that holds its length is initialized at some position, p, and then the array’s data are stored starting at position keccack(p), as shown in Fig.~\ref{fig:arrays}. Solidity’s strings 
and bytes are special dynamic arrays with a somewhat nuanced behavior. As long as they are less than 31 bytes, their data are stored in the same position, p, as their length. When this threshold is exceeded the rules of dynamic arrays apply. 

\paragraph{Storing Data in Event-logs}

Every transaction receipt in the Ethereum blockchain contains log entries (logs), which are indexable checkpoints in EVM code execution \cite{b19}. In the context of Solidity, EVM’s logging operations are facilitated by the use of events. When emitting an event inside a SC, its parameters are stored in logs. Mostly, they are used as a way of triggering a change in an application’s front end or returning a value from a function \cite{b23}. Nonetheless, they can also be used as a means of cheaper data storage. It is noteworthy that the data stored in logs are not accessible from SCs, therefore logs cannot replace storage.

In general, logs consist of the caller’s address, a series of topics and some bytes of data. A bloom filter is utilized to efficiently search for certain logs, based on the caller’s address and topics. When declaring an event, one can specify the keyword indexed for up to three parameters. This will result in a respective number of topics in the produced logs, allowing for a more efficient query, known as filtering. By default, both indexed and non-indexed events store the hash of their signature, which is comprised of information that can later be found in the contract’s ABI and used for filtering, as the first topic. A developer can avoid this by declaring the event as Anonymous, making it less expensive to call (see Table~\ref{table:event_cost}) but more difficult to retrieve.

\begin{table}[]
\caption{Generalized cost model for different types of events}
\label{table:event_cost}
\resizebox{9cm}{!}{%
\begin{tabular}{@{}cccc@{}}
\toprule
 & \textbf{Indexed} & \textbf{Non-Indexed} & \textbf{Anonymous} \\ \midrule
\textbf{Topic{[}0{]}} & Signature Hash (375 gas) & Signature Hash (375 gas) & - \\
\textbf{Topic{[}1{]}} & Indexed Parameter (375 gas) & - & - \\
\textbf{Log Data} & Data (data.len*8 gas) & Data (data.len*8 gas) & Data (data.len*8 gas) \\ \bottomrule
\end{tabular}%
}
\end{table}

\paragraph{Storing Data in Transaction Payload}
In Ethereum there are two types of transactions, those leading to message calls and those which result in contract creation. A message call is the act of passing some value and arbitrary binary data between accounts, either of which can be null. If the target account is a Contract Account (CA), its code is executed with the data part of the message, also termed transaction payload, as input. It is notable that transactions between Externally Owned Accounts (EOAs), may as well contain non-empty payload, transfer no value or even be self-transactions, i.e., the transaction's target is the sender.

As a rule, all transactions, including any payload, are permanently stored in the blockchain. Hence, sending data through a transaction turns out to be an alternative storage method. Considering that 16 gas is charged for every non-zero byte of a transaction’s data and 4 gas for every zero byte, this technique seems appealing. Though, one cannot ignore the fact that such data are not available inside the SC and retrieving them requires storing the corresponding transaction hash, probably in an external database as proposed here \cite{b26}. Apart from that, a transaction payload is, at its core, a hex byte array, which implies that for more complex data types, additional logic should be implemented on the client side.

\paragraph{Storing Data in Unused Function Parameters}

\begin{figure}[htbp]
\centerline{\includegraphics[width=7cm]{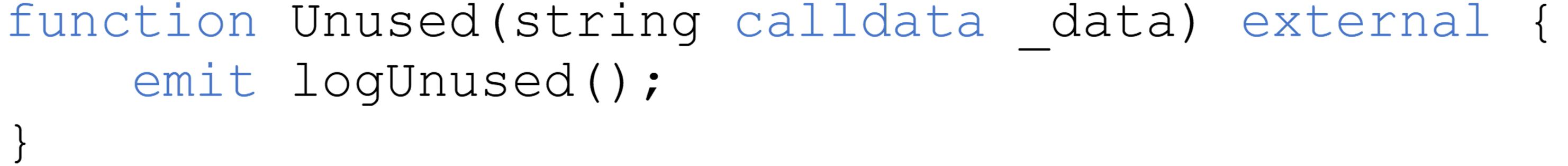}}
\caption{Basic example of unused function parameters.}
\label{fig:un}
\end{figure}

With this term we refer to input data that are passed as function parameters but are no further used within function’s body, as seen in Fig.~\ref{fig:un}. This process is slightly more expensive than the previous one, as it is mandatory for function arguments to be ABI encoded, resulting in a longer payload array. However, it provides better functionality since the underlying tools for encoding and decoding such data are already developed, allowing the use of complex data types.

Moreover, this method being a function call, permits the execution of additional logic during the same transaction, as opposed to the aforementioned method. For instance, an event could be emitted inside the function, providing a way of future-tracking the respective transaction and retrieving its data, eliminating the need for storing the transaction hash externally.

\subsection{Distributed File Systems}
In peer-to-peer distributed file systems, participating nodes share their resources to store and access all kinds of content. Nowadays, IPFS and Swarm are two of the most promising distributed file systems. Both have implemented, evolved, and connected widely researched techniques found in the domain of distributed systems. They aim to be platforms where users can trust the content they receive, without trusting the peers that provide it.

Content is split into chunks before being uploaded to either of these platforms. A chunk is a piece of data that represents the basic unit of storage and retrieval in IPFS and Swarm. Each chunk is uniquely referenced using its cryptographic hash to achieve content addressing, as opposed to the widely used location addressing. This allows for quick verification of received data and enables deduplication. In Swarm, a Binary Merkle Tree (BMT) \cite{b22} hash algorithm based on keccak256 is used to generate the identifiers of the chunks. The developers of IPFS approached this matter differently to provide more flexibility to end users. They introduced CID \cite{b16}: a self-describing upgradable content-addressed identifier, which consists of the hash digest of the content and some metadata as a prefix. From the metadata, the CID version and its encoding, as well as the hashing function used and the content’s type can be determined.

In order to \emph{``group''} chunks of the same data together, both platforms implement some sort of tree data structure. The root hash of such trees refers to the data as a whole. In IPFS a Merkle Directed Acyclic Graph (DAG) \cite{b24}, a generalized construction of a Merkle Tree, is used. It allows nodes to have multiple parents, does not need to be balanced, and non-leaf nodes can contain data. Swarm once more embraced the use of Merkle Trees in which intermediate nodes contain links to leaf nodes that hold the chunks. In the case of uploading folders, IPFS recursively constructs a Merkle DAG whereas Swarm utilizes a Trie \cite{b22} data structure. Both these schemes allow for Unix-like path resolving.

Another fundamental difference between IPFS and Swarm lies in what content participating nodes in each network store. In the former, a node holds content depending on its interests, while in the latter a  \emph{``cloud''} is formed with every node storing arbitrary chunks of data originating from other nodes. In both cases, nodes also cache content they access, until it is \emph{``garbage collected''}. This mechanism is crucial in ensuring that distribution of popular content scales automatically and retrieval latency is decreased. Swarm takes this one step further by implementing an opportunistic cache model: intermediate nodes cache all chunks they forward.

Even though we commented on some basic differences of IPFS and Swarm, there are much more that can’t be reviewed in detail in this work. However, we ought to briefly mention some of them. In IPFS a DHT \cite{b24} is used to map CIDs to seeders that can serve that content. In Swarm, a similar model called DISC \cite{b22} was developed, in which content (value) is directly stored under a corresponding identifier (key), reducing the retrieval process to one routing request. Besides that, in both platforms data are public by default. However, Swarm offers the choice of encrypting data before an upload. Finally, Swarm has a built-in incentive system based on Ethereum, while IPFS depends on a separate blockchain, Filecoin. 

Due to the cost of storing data in Ethereum, most blockchain-based applications with high storage requirements would need to consider an on-chain/off-chain hybrid architecture. Usually, small data are stored directly in the blockchain while larger data, which are not crucial to the SCs operation, are stored off-chain. To achieve timestamping and ensure that the integrity of data stored off-chain can later be verified, their hashes are recorded in the blockchain. This can be done using any of the methods described earlier.

The fact that both IPFS and Swarm rely on Content Addressing for identifying and retrieving data, simplifies the process of incorporating them in the aforesaid hybrid architecture. Identifiers produced during uploads can be directly stored in Ethereum, eliminating the need for additional logic to hash the data.

\section{Experimental Evaluation}

In this section we describe the experimental environment and provide details about the criteria we chose to focus on, and the steps followed in our experiments. Besides that, we discuss the corresponding results.

Before proceeding further, we should mention an important detail concerning the data we used during our experiments. Solidity uses UTF8 encoding, according to which common ASCII characters are represented by a single byte. Additionally, generating random predefined character sequences is simple and all the methods we examined support this format. To evaluate the cost of storage in Ethereum, for each test case, we recorded 14 measurements regarding various data sizes equally spread across the range of 1B-12KB. However, for those cases which exhibited consistent behavior we present only part of our measurements.

At last, it should be pointed out that in some of the measurements we obtained, except for those related to gas costs, we observed abnormal fluctuations. A possible explanation could be that Geth, IPFS and Swarm were running simultaneously on our machine, utilizing the same resources thus affecting each other’s performance.
\subsection{Experimental environment}

We conducted our experiments on a machine with i7-8700 CPU, 64GB RAM, 2TB SSD running Ubuntu 18.04. All SCs were written and compiled in Solidity 0.7.1 with the optimizer disabled and were deployed on Ropsten Testnet. Geth version 1.10.1 was used as an Ethereum client. For IPFS and Swarm, IPFS-client version 0.7.0 and Swarm Bee Client version 0.5.0 were used, respectively.

\subsection{Criteria}
First and foremost, we emphasized on the gas cost related to storing data in Ethereum. For evaluating the sustainability of using Ethereum as a stand-alone data store, each one of the described methods were implemented and the resulting gas costs were compared. In our attempt to use IPFS and Swarm supplementary to Ethereum, the identifiers from uploading data on these platforms were stored in the blockchain and the respective costs were recorded. 

Except for being monetarily viable, DApps need to perform adequately when it comes to retrieving data. That being said, we measured the retrieval time in all the above scenarios. For IPFS and Swarm, both upload and retrieval performances were taken into, aiming to a better overall comparison.

\subsection{Experiments - Storing Data in Ethereum}
\paragraph{Smart Contract Storage}

In order to perform this experiment, we designed a simple SC with the following elements: a public string variable to store our data, a function to modify this string, another one to reset it. In total, we tested three of the most common scenarios.

\begin{itemize}
  \item Storing data after resetting the variable (clean storage).
  \item Updating the variable to a longer string (double the size).
  \item Updating the variable to a new string (same size).
\end{itemize}

\begin{figure}[htbp]
\centerline{\includegraphics[width=9cm]{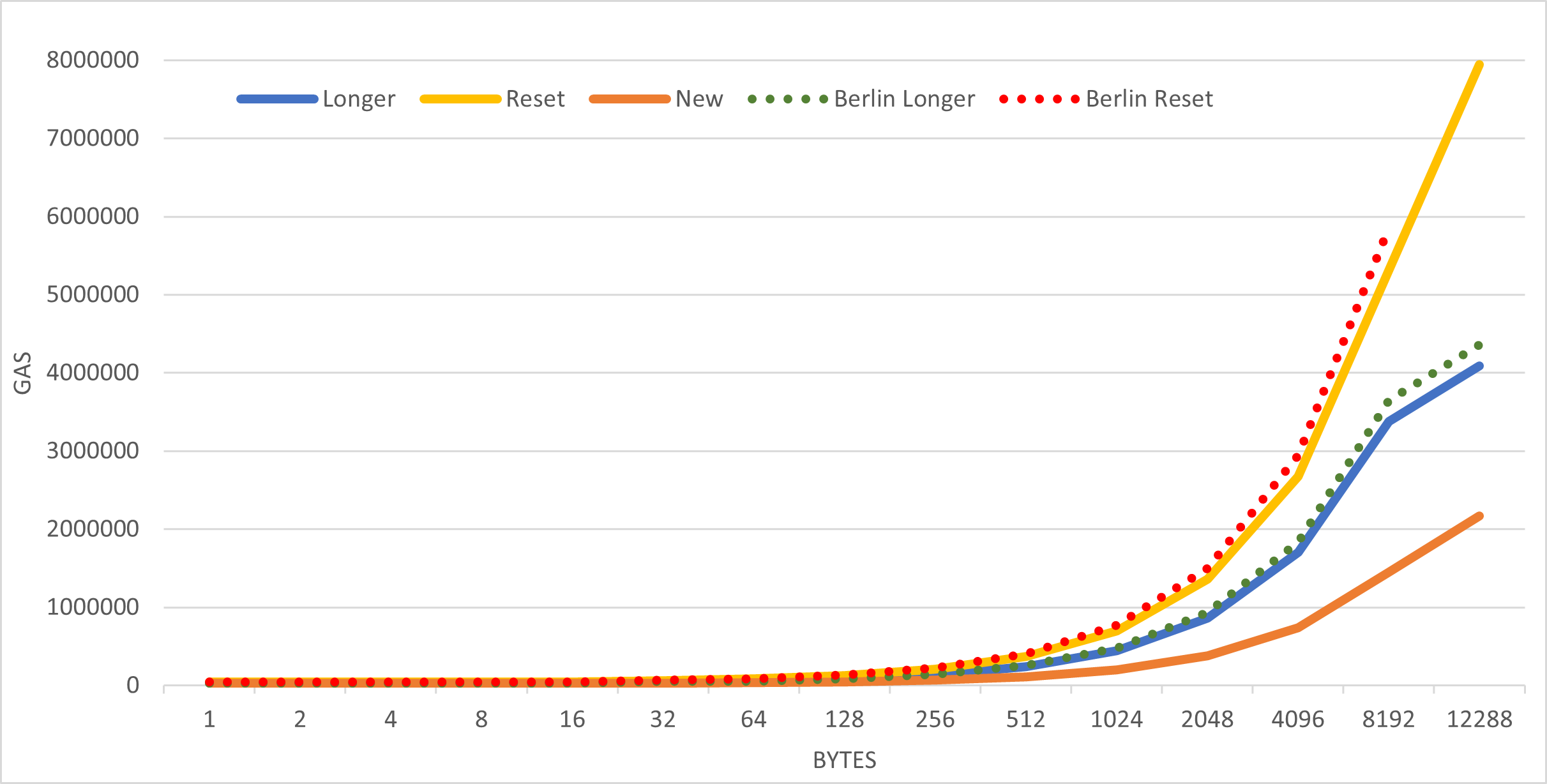}}
\caption{SC storage cost diagram.}
\label{fig:store1}
\end{figure}

\begin{figure}[htbp]
\centerline{\includegraphics[width=9cm]{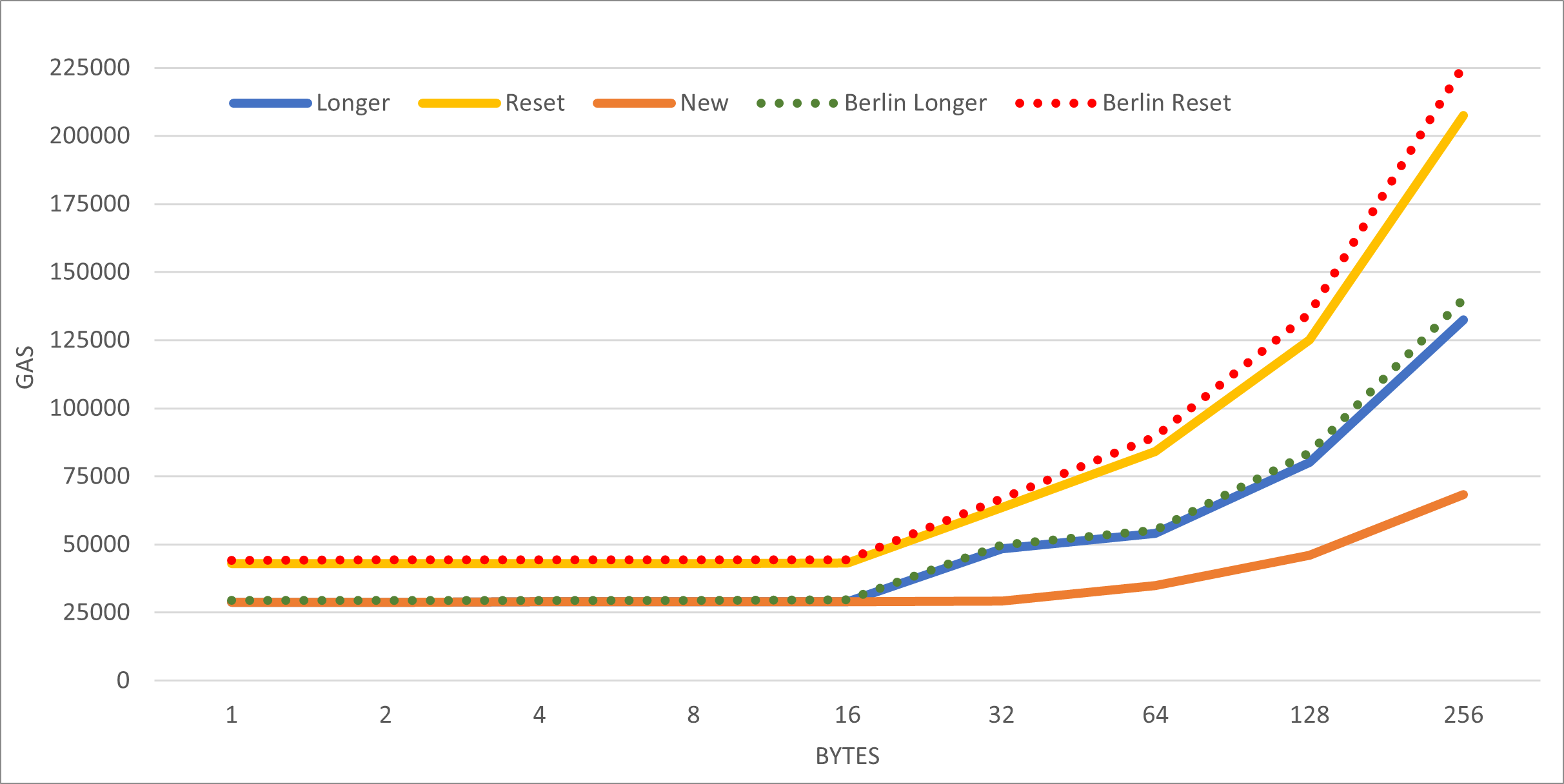}}
\caption{Zoomed in view of the diagram in Fig.~\ref{fig:store1}.}
\label{fig:store2}
\end{figure}

The baseline of 21000 gas for every transaction and the gas paid for its payload are equal in all three cases. So, the number of slots that must be overwritten or initialized during each test mostly accounts for the difference in the respective costs depicted in Fig.~\ref{fig:store1} and Fig.~\ref{fig:store2}, which was included to present the costs for small data. In first case, \(\ceil*{x/32}\) slots for storing the data and one for storing the string's size \(x\) are initialized each time, costing \(20000*num\_slots\) gas. In the last one, 5000 gas is deducted for updating every one of the \(\ceil*{x/32}\) slots. Note that the slot holding the string’s length is not updated, as the sting’s size does not change. Case 2 is a combination of the others and so is the cost calculation method.

Not so long after we executed our experiments, Ethereum Foundation announced a new hard fork named Berlin. Among others, EIP-2929, included in Berlin fork, reforms SLOAD and SSTORE gas metering. We only present the changes that were found to affect our test cases. As reference, the corresponding Geth’s implementation was used \cite{b17}.

\vspace{0.2cm}
\noindent
\textsc{sload} cost modification:
   \begin{itemize}
     \item  \textsc{warm\_sload}: charge 100 gas if slot has already been accessed in current execution context.
     \item  \textsc{cold\_sload}: charge 2100 gas if slot has not been accessed in current execution context.
   \end{itemize}
   
\vspace{0.2cm}
\noindent
 \textsc{sstore} cost modification:
   \begin{itemize}
     \item  Initialize a slot:
     \begin{itemize}
        \item If slot is not already accessed: charge \textsc{cold\_sload} + \textsc{sstore} = 2100 + 20000 = 22100
        \item If slot is already accessed: charge \textsc{sstore} = 20000
     \end{itemize}
   \end{itemize}
   
   \begin{itemize}
     \item  Update a slot: 
     \begin{itemize}
        \item If slot is not already accessed: charge \textsc{cold\_sload} + \textsc{(sstore\_reset - cold\_sload)} = 5000
        \item If slot is already accessed: charge \textsc{(sstore\_reset - cold\_sload)} = 5000 – 2100 = 2900
     \end{itemize}
   \end{itemize}
   
   \begin{itemize}
     \item  Overwriting a slot with the same value (no-op):
     \begin{itemize}
        \item If slot is not already accessed: charge \textsc{(cold\_sload + warm\_sload)} = 2100 + 100 = 2200
        \item If slot is already accessed: charge \textsc{warm\_sload} = 100 
     \end{itemize}
   \end{itemize}
   
Considering the alternated cost model, we re-executed all test cases to examine Ethereum’s actual behavior after the fork. Overall, all transactions of this experiment had a higher cost, especially those regarding our first test case. Actually, the overhead of 2100 for every slot initialization prevented us from storing 12KB because block’s gas limit of 8.000.000 was exceeded. Regarding the third test case, all transactions executed after the fork proved to cost 600 more. By debugging those, we discovered that this difference is caused by the SLOAD gas metering modification. The two SLOAD operations performed on the same slot within the transaction were charged for 2100 and 100 respectively, whereas before they would cost 800 each.

\paragraph{Event-logs}
Prior to any further analysis, we ought to mention that in every event declared throughout the course of this paper’s experiments, an individual counter is included as one of the parameters. Those counters are used as identifiers (ids) and are incremented after each corresponding event is called. This choice was made to simplify the retrieval process. However, the use of the counters entails the execution of an SLOAD operation, which gets their values in order to pass them as parameters to the events and also an SLOAD and SSTORE operation which are needed for incrementing 
them, imposing an overhead of 6600 gas in every transaction. Based on this, it is clear why there is a fairly high cost even in the case of logging one byte.

By repeating the current experiment after the Berlin fork, we observed that the cost of using the aforesaid counters is reduced, namely 1500 gas less in every transaction. In short, the first SLOAD is considered a \emph{``cold''} one whereas the second a \emph{``warm''} one, altogether costing 2200. Also, as the counter’s slot is already accessed, the respective SSTORE costs 2900 instead of 5000, resulting in a total of 5100.

\begin{table}[]
\caption{Structure of events and logs included in the experiments}
\label{table:event}
\resizebox{9cm}{!}{%
\begin{tabular}{@{}cccc@{}}
\toprule
 & \textbf{Indexed} & \textbf{Non-Indexed} & \textbf{Anonymous Indexed} \\ \midrule
\textbf{Declaration} & \begin{tabular}[c]{@{}c@{}}event\_name(uint indexed id, \\ string data)\end{tabular} & \begin{tabular}[c]{@{}c@{}}event\_name(uint id, \\ string data)\end{tabular} & \begin{tabular}[c]{@{}c@{}}event\_name(uint indexed id, \\ string data) anonymous\end{tabular} \\
\textbf{Topic{[}0{]}} & keccak(event\_signature) & keccak(event\_signature) & - \\
\textbf{Topic{[}1{]}} & abi\_encode(id) & - & abi\_encode(id) \\
\textbf{Log Data} & abi\_encode(data) & abi\_encode(id, data) & abi\_encode(data) \\ \bottomrule
\end{tabular}%
}
\end{table}

\begin{figure}[htbp]
\centerline{\includegraphics[width=9cm]{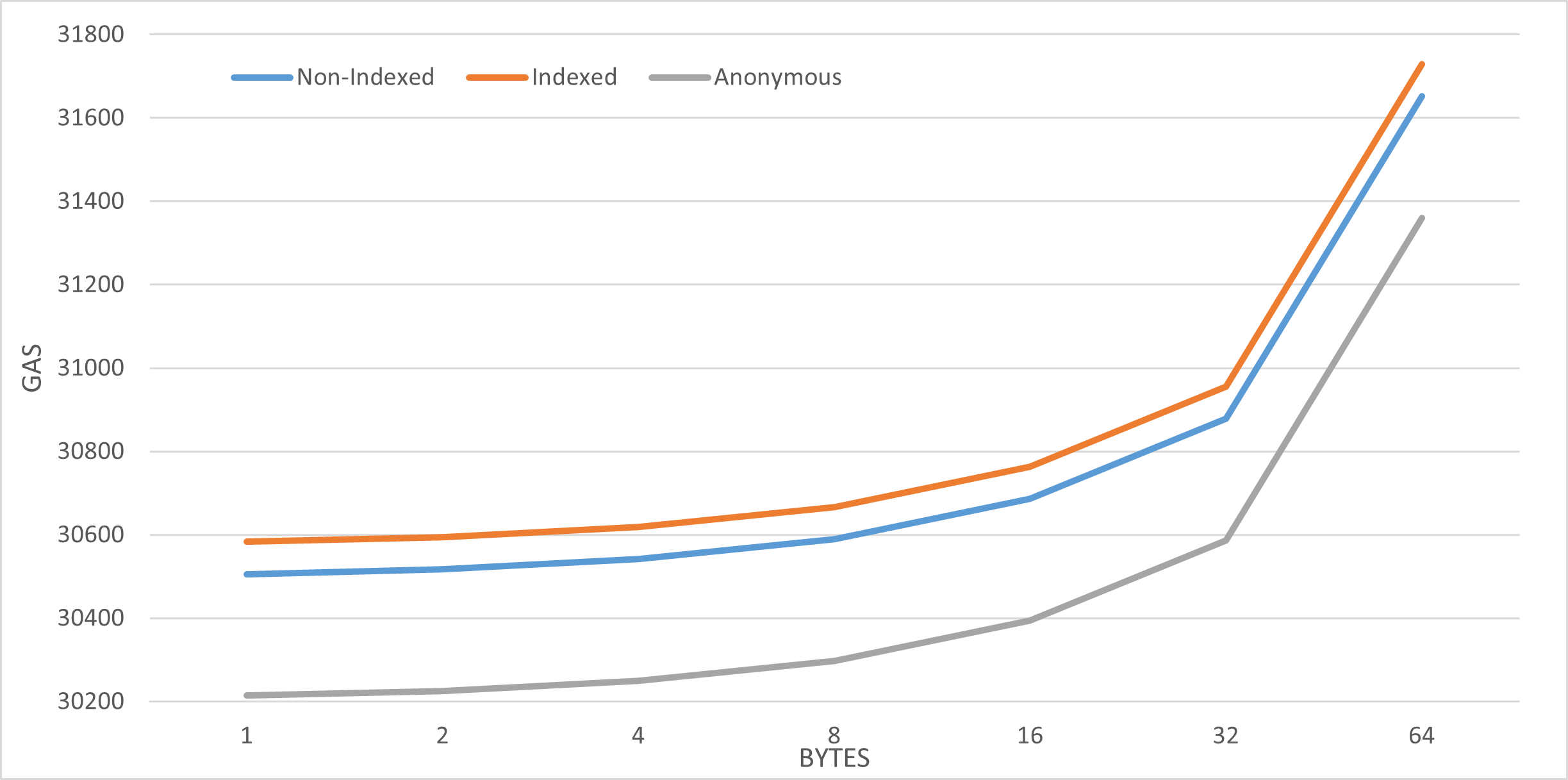}}
\caption{Event-logs storage cost diagram.}
\label{fig:logs}
\end{figure}

As we can observe in Fig.~\ref{fig:logs} the resulting costs from calling each event, exhibit relatively small divergence. Through \cite{b19} we know that each topic costs 375 gas. Respectively, for each byte of data recorded in the data field of the logs there is a charge of 8 gas. Therefore, the number of topics generated by the event call in each of the cases we examined, as well as the amount of data recorded in the log data field, are key factors in the cost difference between them. In our case, both of these are determined by the way the events are declared, as shown in Table~\ref{table:event}. Additionally, since the number of topics is the same in each test, the input data are identical in all three events and the code of the contract is immutable by default, the constant distance maintained by the 3 graphs is justified.

Even though Anonymous events are the least expensive option, they should be used frugally. The fact that they don’t register their signature as a topic, prohibits them from being uniquely referenced, perplexing their retrieval. Including indexed parameters to such events can assist with that.

\paragraph{Transaction Payload}

The first step for this experiment was to encode the data in hex format. Then, we included the resulting hex byte array in the data field of a transaction object, signed it and sent it. Note that no ether was transferred during any of the transactions. We tried two different ways of implementing this method:

\begin{itemize}
  \item Sending the transaction between EOAs
  \item Sending the transaction from an EOA to a CA
\end{itemize}

As expected, the measurements confirm that the use of transactions as a data store is the most cost-efficient solution of all. As a matter of fact, this can be considered the minimum cost for storing data in Ethereum. That is, because the fee paid for a transaction’s payload is essentially included in every available option, e.g for executing a transaction with some data as input, these data must first be sent as payload to the contract.

\begin{figure}[htbp]
\centerline{\includegraphics[width=9cm]{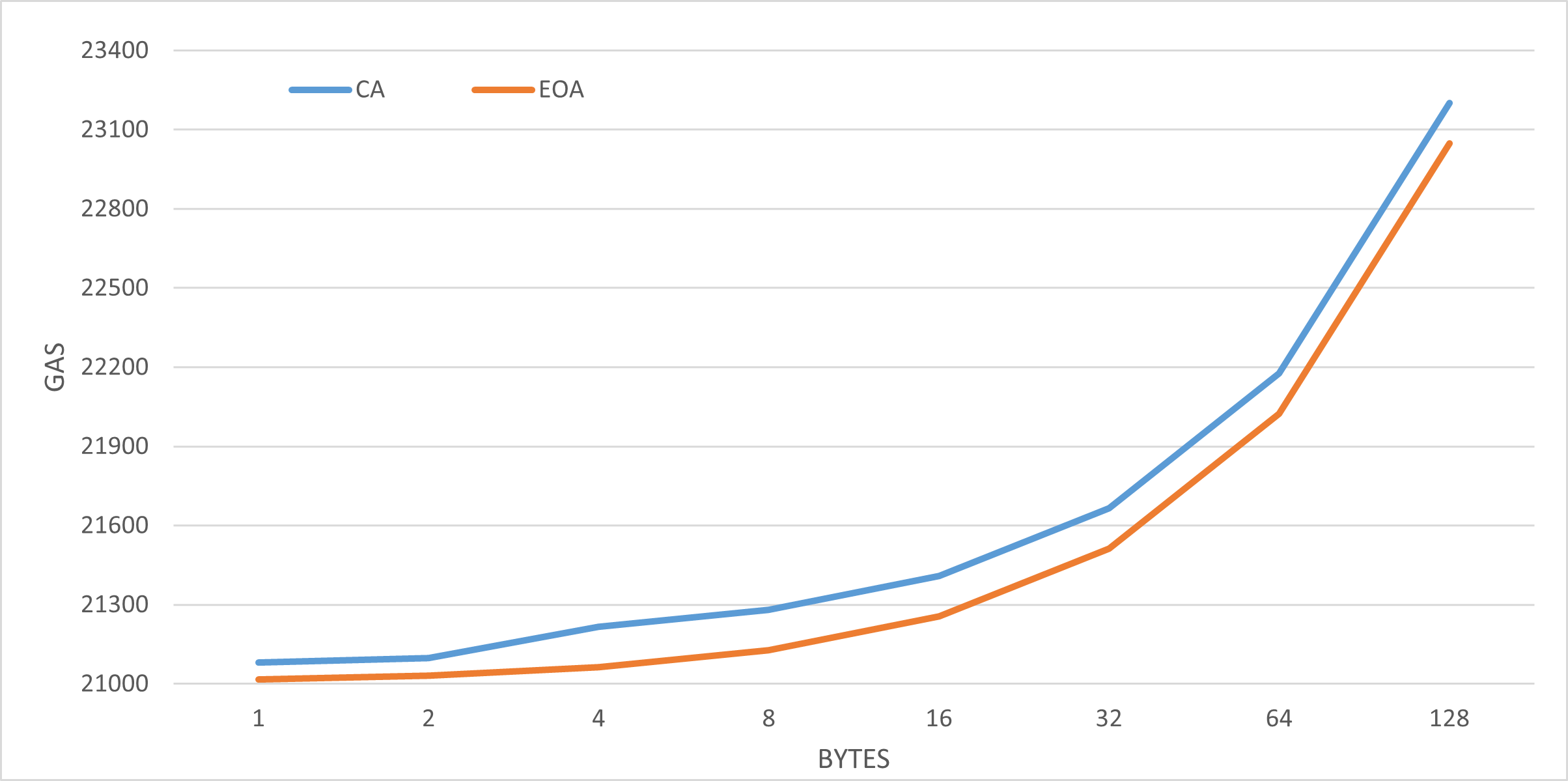}}
\caption{Transaction payload storage cost diagram.}
\label{fig:tx}
\end{figure}

For the second case we tested, a fallback function in the targeted SC is required, otherwise the transactions will fail. In short, fallback is a special function that gets executed if none of the function selectors match the first four bytes of the transaction payload, or if the latter is empty and no Receive Ether function is defined. Additionally, it cannot have parameters nor return anything.

The difference between the amount of gas spent in each test case, depicted in Fig.~\ref{fig:tx}, stems from the fact that in the second one some contract code needs to be executed. By inspecting the bytecode of the corresponding contract, we managed to further analyze this difference. After verifying that no ether was sent with the transaction, a series of opcodes is executed to check if the payload is less than four bytes. If that is true, EVM jumps to fallback’s execution. The accumulated cost resulting from the EVM execution, up to that point, is 65 gas greater than that of test Case 1. If the aforementioned condition is false, the first four bytes of the payload are loaded and processed, costing 12 gas, and they are compared against every function selector of our contract’s three functions. Eventually no match is found and the execution jumps to the fallback function, which costs 10 gas. The code blocks in which the comparisons are conducted consist of the opcodes \textsc{\{dup1, push4, eq, push2, *jumpi\}}, which altogether cost 22 gas. Taking into account that no match will be found, all comparisons are executed, costing 66. In total, the cost for storing \(data\geqslant 4\) bytes in a transaction by exploiting the fallback function, is 153 gas greater than the other test case.


By inspecting the execution of the contract at bytecode level we managed to verify what to our knowledge was first reported, but not explained, in \cite{b18}. Simply put, function selectors, which depend on function names, are compared to the first four bytes of a transaction’s payload in hex-ascending order. Hence, frequently used functions should be named in an appropriate manner to avoid numerous comparisons, i.e., save gas. Moreover, we believe it is necessary to point out that for
a contract with a large number of functions, solidity optimizes
the comparison flow by performing a pre-comparison to check
whether the given function selector is or is not above a certain
threshold.

\paragraph{Unused Function Parameters}
This experiment was performed to examine the cost of storing data in a transaction’s payload while exploiting Solidity’s built-in ABI interface, which enables the use of more complex data types. To achieve this we used a function with a single parameter that was no further processed. We also decided to improve the functionality of this method by emitting an event, with just one indexed parameter, inside the function in order to later track the hash of the transaction. This can also be done when utilizing a fallback function.

\begin{figure}[htbp]
\centerline{\includegraphics[width=9cm]{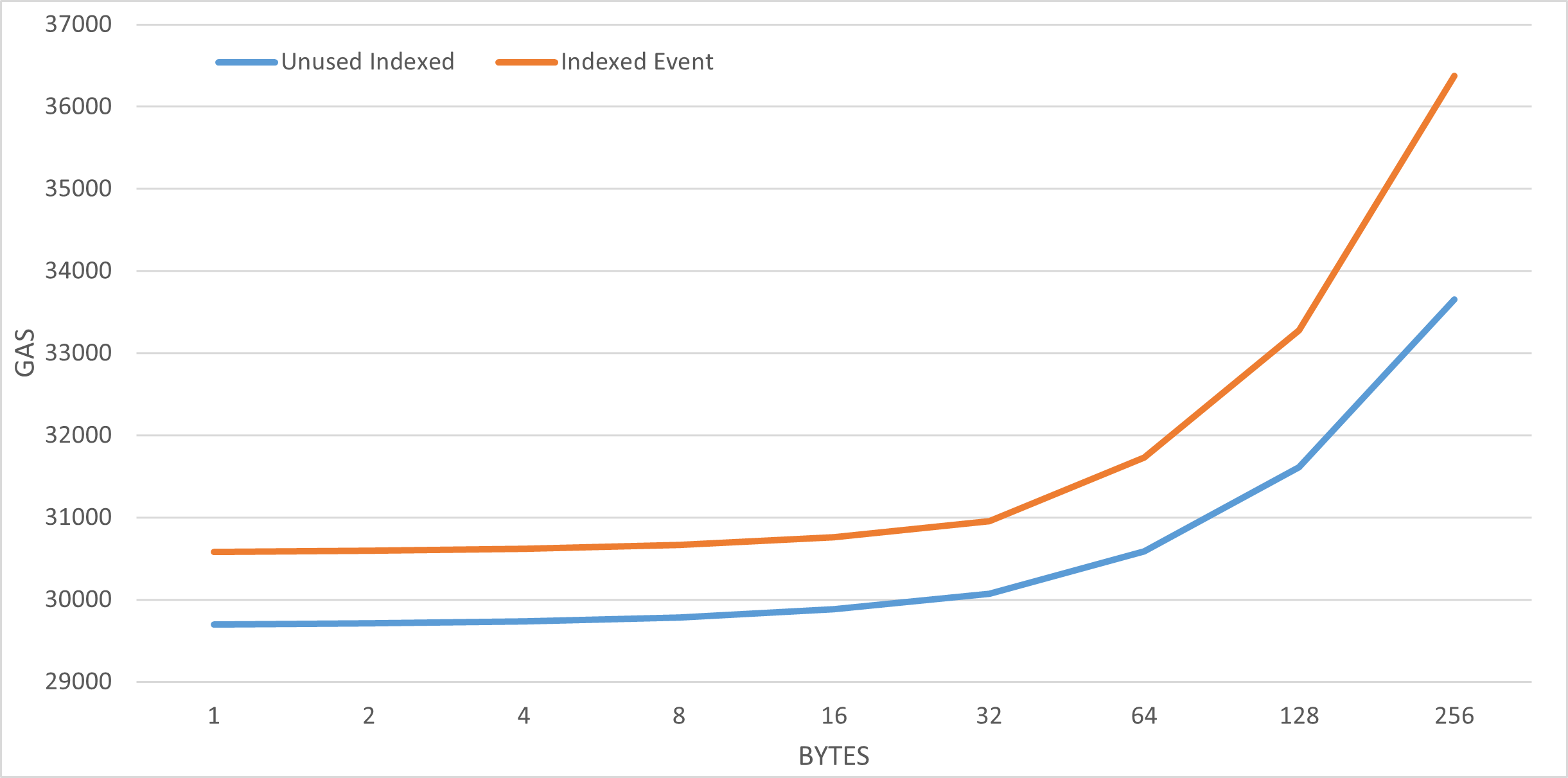}}
\caption{Unused function parameter and event-logs storage cost diagram.}
\label{fig:unused}
\end{figure}

To evaluate the recorded costs, we used the case of indexed events as a reference point. It is obvious from Fig.~\ref{fig:unused} that as input data grow in size, the current approach gets even cheaper than the other. The fact that in this case, no data are recorded in the data part of the produced log, accounts for the difference.

It is noteworthy, that log data are also ABI encoded, resulting in longer log arrays. For example, a Boolean which is one byte, after encoding takes up 32 bytes and costs \(\ 8*32 = 256\) gas to be stored in a log. So, in a scenario in which more unused parameters are utilized, the current approach is even less expensive. To confirm this, we conducted a complementary experiment, locally in Ganache. A contract with two functions carrying seven parameters each was deployed. Function 1 emitted an event with all parameters as input, whereas Function 2 emitted an event with just the first parameter. After executing both functions with the same arguments, we concluded that the utilization of unused function parameters can be quite effective, as Function 2 cost nearly 7000 less gas.

At this point, we would like to discuss some possible drawbacks regarding this alternative storage method. First, by testing we discovered that Solidity imposes a restriction on the number of parameters a function can bear, namely sixteen. If this threshold is exceeded, a \emph{``stack too deep''} error is thrown. Additionally, defining a function with unused function parameters causes a warning at compilation. Thus, it might be disallowed in future Solidity versions.

\subsection{Experiments - Data Retrieval in Ethereum}

In this paragraph, based on our experimental measurements, we present an overall comparison of the time needed to retrieve data from the Ethereum blockchain. Each measurement refers to the retrieval time of the data that were stored during the previous experiments. We noticed that the results were quite similar for the different data sizes (1B – 12KB), so, in Table~\ref{tab:retrieve} we recorded the average of these results. Also, we should mention that retrieving data stored in unused function parameters or Anonymous events is a two-step process. Likewise, using events without indexed parameters implies that additional logic should be implemented to find the desired data among the returned events. In any case, these delays were found to be insignificant compared to the time needed to retrieve the respective events. Thus, all measurements presented in Table~\ref{tab:retrieve} include any necessary additional step.

\begin{table}[]
\caption{Retrieval time in Ethereum (ms)}
\label{tab:retrieve}
\resizebox{9cm}{!}{%
\begin{tabular}{@{}ccc@{}}
\toprule
 & \multicolumn{2}{c}{\textbf{Retrieval Time}} \\ \midrule
\textbf{Storage} & \multicolumn{2}{c}{6.42} \\
\textbf{Transaction} & \multicolumn{2}{c}{6.47} \\ \midrule
\textbf{} & \textbf{fromBlock: 0} & \textbf{\begin{tabular}[c]{@{}c@{}}fromBlock: con\_creation\end{tabular}} \\ \midrule
\textbf{Non-Indexed Event} & 390.21 & 84.45 \\
\textbf{Indexed Event} & 144.83 & 30.47 \\
\textbf{Anonymous Indexed Event} & 707.45 & 37.77 \\
\textbf{Unused Non-Indexed} & 289.19 & 68.48 \\
\textbf{Unused Indexed} & 167.17 & 31.45 \\\bottomrule
\end{tabular}%
}
\end{table}

Considering the measurements presented in Table~\ref{tab:retrieve}, it is obvious that both retrieving data from contract’s storage or retrieving a transaction and extracting its data, require substantially less time than that of the other test cases. That is because these particular retrieval processes are ultimately single database queries that target either the Storage Trie or a Transaction Trie. 

On the other hand, the process of retrieving events involves the utilization of Bloom filters and therefore is more complex and leads to higher latency. Though, through an appropriate interface, a developer can specify the block from which the client should start searching for the requested event(s). This can greatly reduce the number of database queries and consequently result in faster retrieval. Indeed, we confirmed that searching from the beginning of the blockchain as opposed to specifying a more recent starting point requires significantly more time. In our experiments, as a starting block, we set the ones that our contracts were deployed in, because no events could have been emitted before these. Besides that, we observe that the use of indexed parameters in an event can further improve the results.

Driven by the fact that most operating systems utilize free RAM memory for caching, we decided to assess how the system’s cache may influence retrieval performance. So, we repeated our experiments but dropped the page cache between each run, as it is proposed in \cite{b28}. This dramatically delayed the retrieval of data from logs when the starting block was set to zero (i.e., beginning of the chain). When the search was conducted with a more recent starting point, the latency was tolerable. Due to limited space, we will present these measurements in the long version of this work.

To sum up, retrieving logs is time consuming and imposes an overhead in most of the methods we proposed. However, one can make this compromise considering the cost reduction that these alternatives bring. Apart from that, if managing an external database for storing the transaction hashes is not discouraging, data could be stored in a transaction’s payload or even in unused function parameters, without emitting an event, which imposes an extra cost. By doing so, the cost as well as the retrieval time are reduced substantially. 

\subsection{Experiments - IPFS \& SWARM}

The aim of this experiment was to examine the use of IPFS and Swarm alongside Ethereum. We stored data blocks of size 4KB – 16MB in these platforms and recorded the resulting identifiers in Ethereum using both SC storage and logs. The cost related to storing these identifiers in Ethereum was examined, but due to space limitation it can’t be presented. In general, it is consistent due to the constant length of the identifiers. However, in IPFS, using a CID v1 which is longer, instead of CID v0, will result in higher gas consumption. This also applies to Swarm when encrypting files because it leads to an identifier double the normal size. 

\begin{figure}[htbp]
\centerline{\includegraphics[width=9cm]{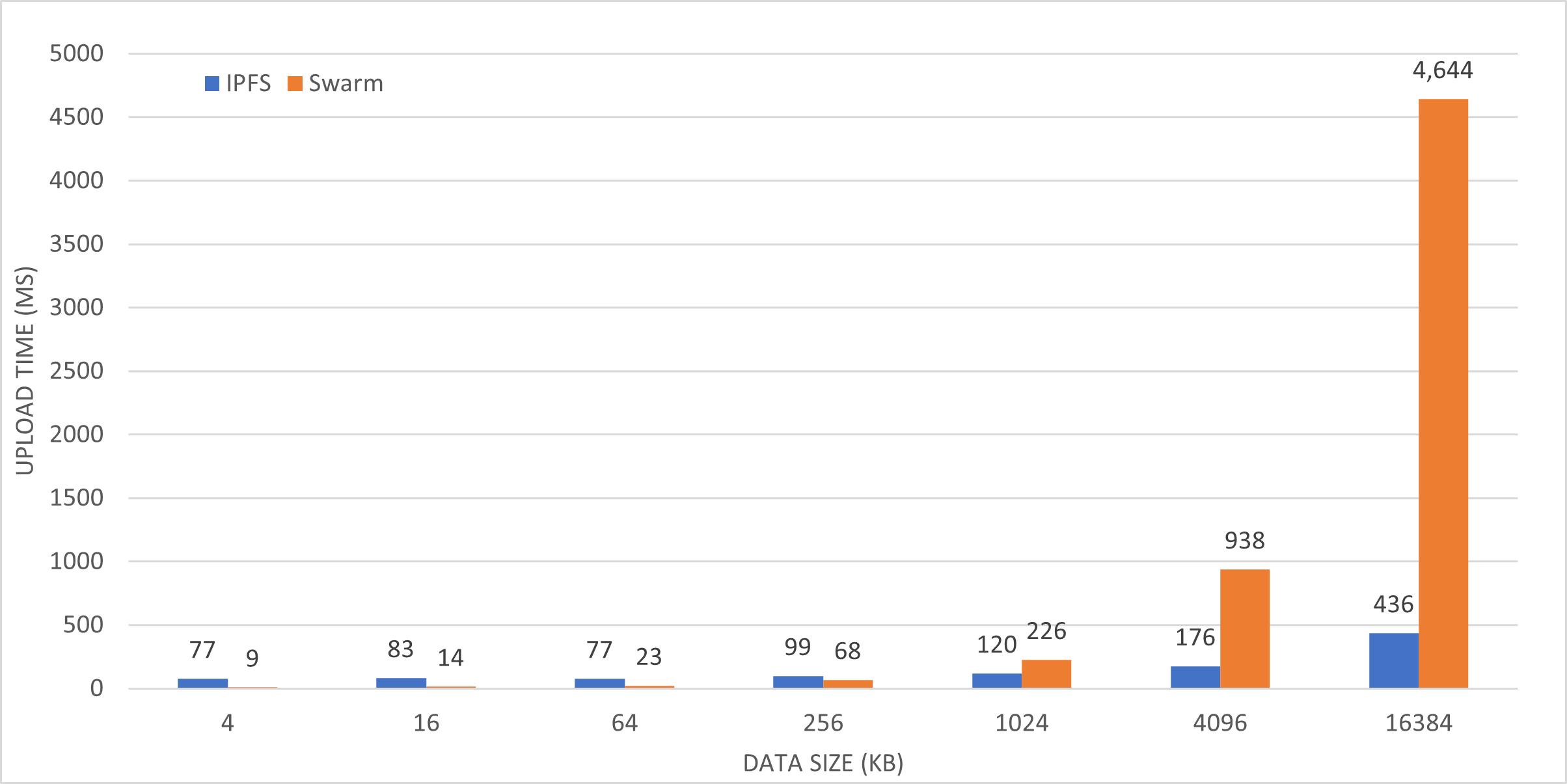}}
\caption{IPFS-Swarm upload latency.}
\label{fig: ipfs_swarm_upload}
\end{figure}

The measurements presented in Fig.~\ref{fig: ipfs_swarm_upload}, indicate that uploading data up to 256KB in Swarm is quite fast. Above this threshold performance deteriorates substantially. We must consider that data are split into chunks which are hashed and organized in a Binary Merkle Tree. Obviously, the number of calculations included in this process is proportional to the size of the data and definitely affects the upload-time.

IPFS behaves similarly, except that the size of the data has smaller impact on its performance. This mostly stems from the fact that chunk’s default size (256KB) in IPFS is 64 times the size of a Swarm chunk (4KB), rendering the process of splitting the data and organizing the chunks in a Merkle-Dag faster. Accordingly, the quite stable measurements for data up to 256KB are justified, as they fit in one chunk. To substantiate this claim, we uploaded data on IPFS with the chunker option set to 4KB instead of 256KB. In all tests, the upload latency increased significantly. In fact, Swarm proved to be far more efficient in this scenario. On top of that, the different algorithms and data structures used by each platform to split, store and link data, contribute to the performance gap between them.

\begin{figure}[htbp]
\centerline{\includegraphics[width=9cm]{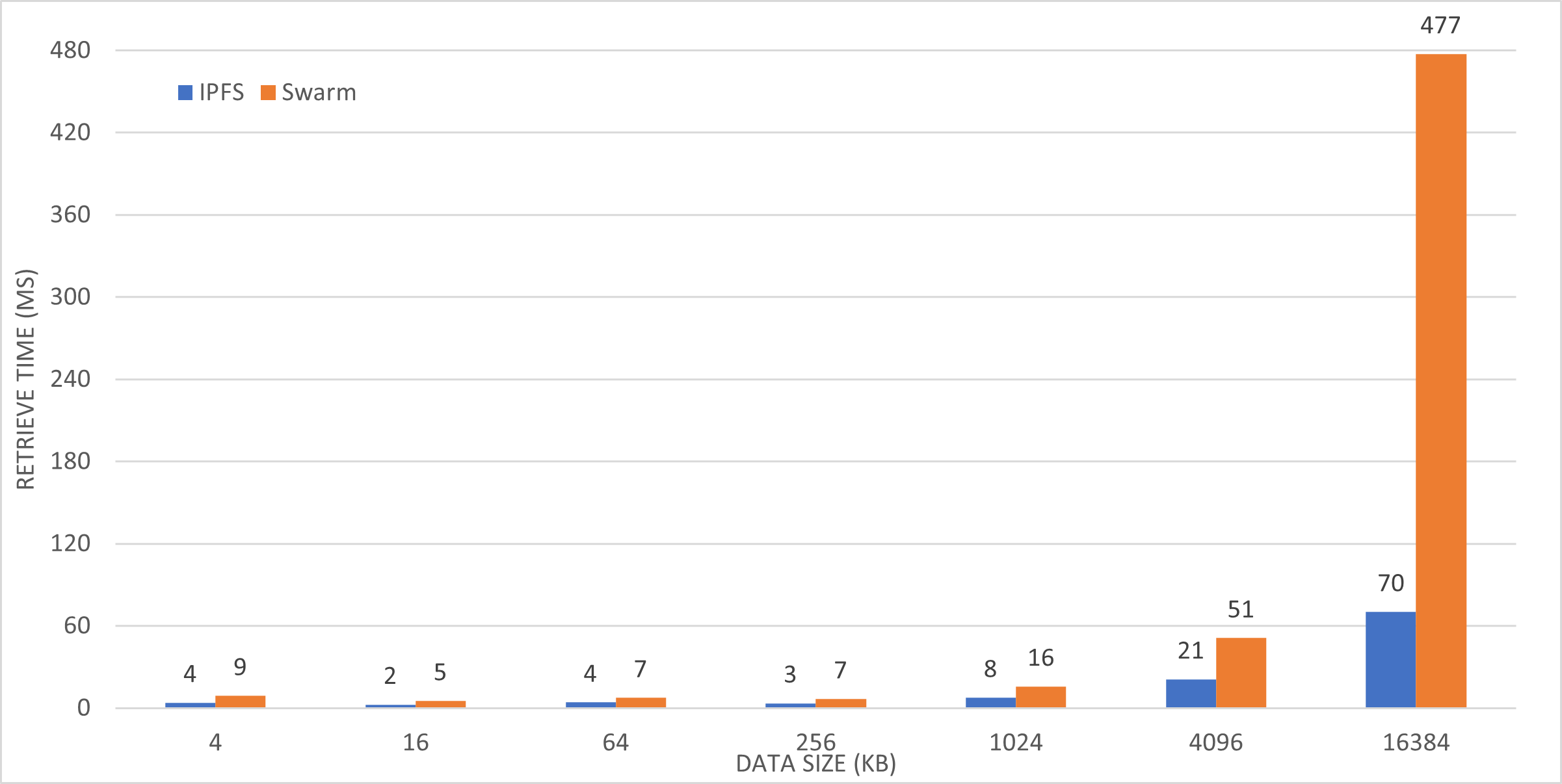}}
\caption{IPFS-Swarm retrieval latency.}
\label{fig: ipfs_swarm_retrieve}
\end{figure}

As we can observe from Fig.~\ref{fig: ipfs_swarm_retrieve}, Swarm lacks in performance when it comes to retrieving data from local storage. That is due to the fact that all chunks split during the upload process must be individually fetched in order to reconstruct the original data. A large number of chunks implies a complex Merkle-Tree or Merkle-DAG. Thus, longer paths have to be resolved until each chunk is found, leading to greater retrieval time. It is clear that once again, the chunk-size utilized within each platform’s architecture is a determining factor for their performance.

We ought to clarify that by conducting these experiments we meant to compare the performance of IPFS and Swarm, regarding local upload and retrieval. The results are nothing but representative of how these platforms behave when exchanging data between remote nodes. Besides that, Swarm updated its networking protocol from DevP2P to LibP2P and introduced a new client, which we used during our experiments. The fact that it is in primary stage and improvements are frequently made might render our results obsolete in the near future.

\section{Conclusions}



We evaluated a wide range of data storage options for Ethereum DApps. We examined the cost related to storing data in SC storage, as well as in alternative data stores, like event-logs and unused function parameters, and hybrid approaches based on IPFS or Swarm. Through a comparative study, we identified benefits and drawbacks of each approach.

The results of our work confirm that there exists a variety of options for Ethereum DApps to store data, which significantly differ with each other with respect to the cost, retrieval latency and the associated implementation complexity. Consequently, it is of critical importance for DApp designers and developers to choose the appropriate data management scheme for their Ethereum applications. In this context, we hope that our observations can be used as a guide in the search for proper data management methods.

Due to limited space, a large number of experiments and measurements have been omitted from the present work and will be included in its long version. Moreover, in future work, we plan to evaluate our methods on real-world use-cases. Furthermore, we will perform experiments to assess the performance of IPFS and Swarm regarding the retrieval of data from remote nodes. Finally, applying formal methods on the presented data management approaches similar to the ones used in~\cite{b4} for SC, might also be an interesting direction.


\begin{thebibliography}{00}
\bibitem{b20}V. Buterin, “A next-generation smart contract and decentralized application platform,” white paper, vol. 3, no. 37, 2014
\bibitem{b24} J. Benet, “IPFS - Content Addressed, Versioned, P2P File System,” arXiv:1407.3561 [cs], Jul. 2014, Accessed: Apr. 13, 2021. [Online]. Available: http://arxiv.org/abs/1407.3561.
\bibitem{b22}V. Trón, The Book of Swarm: storage and communication infrastructure for self-sovereign digital society back-end stack for the decentralised web, V1.0 pre-Release 7. 2020.
\bibitem{b6} N. Grech, M. Kong, A. Jurisevic, L. Brent, B. Scholz, and Y. Smaragdakis, “MadMax: analyzing the out-of-gas world of smart contracts,” Commun. ACM, vol. 63, no. 10, pp. 87–95, Sep. 2020.
\bibitem{b27} C. Signer, “Gas cost analysis for ethereum smart contracts,” Master’s Thesis, ETH Zurich, Department of Computer Science, 2018.
\bibitem{b1} Y. Kurt Peker, X. Rodriguez, J. Ericsson, S. J. Lee, and A. J. Perez, “A Cost Analysis of Internet of Things Sensor Data Storage on blockchain via Smart Contracts,” Electronics, vol. 9, no. 2, Art. no. 2, Feb. 2020
\bibitem{b10} T. Chen, X. Li, X. Luo, and X. Zhang, “Under-optimized smart contracts devour your money,” in 2017 IEEE 24th International Conference on Software Analysis, Evolution and Reengineering (SANER), Feb. 2017, pp. 442–446
\bibitem{b4} E. Albert, J. Correas, P. Gordillo, G. Román-Díez, and A. Rubio, “Don’t run on fumes—Parametric gas bounds for smart contracts,” Journal of Systems and Software, vol. 176, p. 110923, Jun. 2021.
\bibitem{b8} B. Yankov, “Storing and Retrieval of Structured Data on blockchain with BlockChi and Ethereum,” vol. 15, no. 1, p. 10, 2019.
\bibitem{b26}C. Xie, Y. Sun, and H. Luo, “Secured Data Storage Scheme Based on Block Chain for Agricultural Products Tracking,” in Proceedings - 2017 3rd International Conference on Big Data Computing and Communications, BigCom 2017, 2017, pp. 45–50
\bibitem{b9} V. K. Calastry Ramesh, “Storing IOT Data Securely in a Private Ethereum blockchain,” UNLV Theses, Dissertations, Professional Papers, and Capstones, May 2019 
\bibitem{b11} J. Shen, Y. Li, Y. Zhou, and X. Wang, “Understanding I/O Performance of IPFS Storage: A Client’s Perspective,” in 2019 IEEE/ACM 27th International Symposium on Quality of Service (IWQoS), Jun. 2019, pp. 1–10
\bibitem{b19}G. Wood, “Ethereum: A secure decentralised generalised transaction ledger,” Ethereum project yellow paper, vol. 151, no. 2014, pp. 1–32, 2014.
\bibitem{b13} “Solidity — Solidity 0.8.3 documentation.”  https://docs.soliditylang.org/en/v0.8.3/ (accessed Apr. 09, 2021).
\bibitem{b23}“A Guide to Events and Logs in Ethereum Smart Contracts,” ConsenSys. https://consensys.net/blog/developers/guide-to-events-and-logs-in-ethereum-smart-contracts/ (accessed Apr. 13, 2021).
\bibitem{b16} multiformats/cid. Multiformats, 2021.
\bibitem{b17} “ethereum/go-ethereum,” GitHub.
 https://github.com/ethereum/go-ethereum (accessed Apr. 10, 2021).
\bibitem{b18} Y.-C. Chen, “[Solidity] How does function name affect gas consumption in smart contract,” Medium, Sep. 11, 2018. https://medium.com/joyso/solidity-how-does-function-name-affect-gas-consumption-in-smart-contract-47d270d8ac92 (accessed Apr. 10, 2021).
\bibitem{b28}D. Perez and B. Livshits, “Broken Metre: Attacking Resource Metering in EVM,” arXiv:1909.07220 [cs], Mar. 2020, Accessed: Apr. 13, 2021. [Online]. Available: http://arxiv.org/abs/1909.07220.

\end{thebibliography}
\end{document}